\documentclass[traditabstract]{aa}
\usepackage{natbib}
\usepackage{amsmath}
\usepackage{graphicx}
\usepackage{wasysym}
\usepackage{txfonts}
\usepackage{xspace}

\newcommand{\inte}{\textsl{INTEGRAL}\xspace}

\newcommand{\xte}{\textsl{RXTE}\xspace}
\newcommand{\chandra}{\textsl{Chandra}\xspace}

\newcommand{\msun}{\ensuremath{\text{M}_{\odot}}}

\newcommand{\nh}{\ensuremath{N_\text{H}}\xspace}

\begin{document}

\title{X-ray variation statistics and wind clumping in Vela~X-1}

\titlerunning{X-ray variation statistics in Vela~X-1}

\author{
 \mbox{Felix F\"urst\inst{1}} \and
  \mbox{Ingo Kreykenbohm\inst{1}} \and
\mbox{Katja Pottschmidt\inst{2,3}} \and
\mbox{J\"orn Wilms\inst{1}} \and
\mbox{Manfred Hanke\inst{1}}\and
\mbox{Richard E. Rothschild\inst{4}} \and
\mbox{Peter Kretschmar\inst{5}} \and
\mbox{Norbert S. Schulz\inst{6}} \and
\mbox{David P. Huenemoerder\inst{6}}\and
\mbox{Dmitry Klochkov\inst{7}} \and
\mbox{R\"udiger Staubert\inst{7}}
}

\authorrunning{F.~F\"urst et al.}

\institute{
 Dr.~Karl Remeis-Sternwarte \& ECAP,  Universit\"at Erlangen-N\"urnberg, Sternwartstr.~7, 96049~Bamberg, Germany
\and CRESST and NASA Goddard Space Flight Center, Astrophysics Science Division, Code 661, Greenbelt, MD 20771, USA
\and Center for Space Science and Technology, University of Maryland Baltimore County, 1000 Hilltop Circle, Baltimore, MD 21250, USA
\and Center for Astrophysics \& Space Sciences, University of California, San
Diego, 9500 Gilman Drive, La Jolla, CA 92093, USA
\and European Space Agency, European Space Astronomy Centre, Villafranca del Castillo, 
P.O.~Box~78, 28691~Villanueva de la Ca\~{n}ada, Madrid,
Spain
\and
Center for Space Research, Massachusetts Institute of Technology, 77 Massachusetts Avenue, Cambridge, MA 02139, USA
\and
Kepler Center for Astro and Particle Physics, Institut f\"ur Astronomie und
Astrophysik,\\
Universit\"at T\"ubingen, Sand~1, 72076~T\"ubingen, Germany
}

\date{Received: --- / Accepted: ---}

  \abstract{ 
We investigate the structure of the wind in the neutron star X-ray binary
system Vela~X-1 by analyzing its flaring behavior. Vela~X-1 shows 
constant flaring, with some flares reaching fluxes of more than 3.0\,Crab between 20--60\,keV for several 100\,seconds, while the average flux is around 250\,mCrab. We analyzed all archival \textit{INTEGRAL} data,
calculating the brightness distribution in the 20--60\,keV band, which, as we show, closely follows a log-normal distribution. Orbital resolved analysis shows that the structure is strongly variable, explainable by shocks and a fluctuating accretion wake. Analysis of \xte ASM data suggests a strong orbital change of \nh. Accreted clump masses derived from the \inte data are on the order of $5\times10^{19}$--$10^{21}$\,g. 
We show that the lightcurve can be described with a model of multiplicative random numbers. In the course of the simulation we calculate the power spectral density of the system in the 20--100\,keV energy band and show that it follows a red-noise power law.
We suggest that a mixture of a clumpy wind, shocks, and turbulence can explain the measured mass distribution.
As the recently discovered class of supergiant fast X-ray transients (SFXT) seems to show the same parameters for
the wind, the link between persistent HMXB like Vela~X-1
and SFXT is further strengthened.
}


\maketitle

\section{Introduction}
The high-mass X-ray binary (HMXB) Vela~X-1 was discovered in 1967 \citep{chodil67a} and
is today one of the best studied objects of its class hosting a neutron star. It is a bright, eclipsing
source, showing strong pulsations with a pulse period of around $283$\,sec 
\citep{mcclintock76a} and an orbital period of 8.9\,days
\citep{vankerkwijk95a}. \citet{quaintrell03a} have shown that the separation
between the neutron star and its optical companion, the B0.5Ib supergiant
HD\,77581, is only
$1.7$\,stellar radii. The neutron star is thus deeply embedded in the
stellar wind of the supergiant, which has a mass loss rate of about $10^{-6}\,\msun\,$yr$^{-1}$
\citep{nagase86a}. Material from the wind is accreted by the neutron star and is
channeled by its strong magnetic field onto the magnetic poles. The
accretion rate leads to an average X-ray luminosity of $\sim$ $4 \times
10^{36}$\,erg\,s$^{-1}$, although the luminosity is strongly variable on all
time-scales, varying up to a factor of at least 20--30 \citep{staubert04a, kreykenbohm08a}. \citet{staubert04a} found short, but very bright flares reaching up to almost 7\,Crab (1000\,counts\,sec$^{-1}$ [cps] in the 20--40\,keV band of \inte ISGRI).

The physical processes leading to this strong variation are unclear, however. Because the
luminosity of the neutron star is proportional to the mass accretion rate, changes in the density or the effective velocity of the stellar wind
directly lead to variations in the X-ray luminosity \citep{bondi44a, davidson73a}. Changes of that kind have shown up in simulations of the system \citep{blondin90a, blondin91a, mauche08a}. Strong density changes have also been proposed to explain varying line strengths in high-resolution X-ray spectra \citep{sako99a, watanabe06a} or were inferred via measurement of emission lines in hot, thin plasma at the same time with fluorescence lines in colder, optical thick plasma \citep{schulz02a, goldstein04a}.

The goal of this paper is to study the variations in the lightcurve and investigate a possible connection between the wind structure and the flaring behavior.
This paper is structured as follows:
in Sect. \ref{sec:obsdat} we present our data and analysis method.
In Sect. \ref{sec:ana_all} we show an orbital phase averaged and orbital phase resolved analysis of the \inte ISGRI
hard X-ray and \xte ASM soft X-ray lightcurve. In order to constrain the physical processes leading to the observed variability we present a numerical method to simulate lightcurves in Sect. \ref{sec:simu}. For this simulation we calculate and model the power spectral density of Vela~X-1. In Sect. \ref{sec:disc} we summarize our results and discuss them in a larger context.

\section{Observations and data} \label{sec:obsdat}
The ``International Gamma-Ray Astrophysics Laboratory'' \citep[\inte,][]{winkler03a} carries three X-ray instruments: JEM-X \citep{lund03a} for the soft X-rays, the spectrometer SPI \citep{verdenne03a} and  the ``Imager on Board the \inte Satellite'' (IBIS) for the hard X- and $\gamma$-rays \citep{ubertini03a}. All these instruments are coded mask instruments \citep[][and references therein]{renaud06a}, allowing imaging of the hard X-ray sky. IBIS has a field of view of almost $30\degr\times30\degr$ and its detector consists of two planes: the bottom layer PICsIT, sensitive in the 200\,keV--10\,MeV range and the top layer ISGRI, which is sensitive between 15\,keV--1\,MeV \citep{lebrun03a}. Because of the coded mask instruments, observations with \inte are usually carried out in a dithering mode, where the pointing of the satellite is shifted slightly every 1800--3600\,sec. Each of these pointings makes up one science window (ScW). During slews between ScWs, taking about 150\,sec, no data are gathered.
\inte has provided a large fraction of the detailed hard X-ray studies of Vela~X-1, producing 
almost 3.6\,Msec of good data between 2003 November and 2006 May, split mainly into three large blocks separated by some hundred days.
All observations of \inte that were public as of 2009 December 01 were used as the data basis for this work, see Table \ref{tab:blocks}.

\begin{table}
\centering
\caption{Overview of the \inte data of Vela X-1}
\begin{tabular}{llll}
\hline\hline
  & Revolutions & MJD & Exposure \\\hline
 Block 1  & 137 -- 141  & 52970.4 -- 52984.9 & 0.92\,Msec \\
 Block 2  & 373 -- 383  & 53678.7 -- 53708.2  & 1.57\,Msec \\
 Block 3  & 433 -- 440  & 53857.1 -- 53879.0 & 1.13\,Msec \\\hline
\end{tabular} 
\label{tab:blocks}
\end{table} 

\section{Statistical analysis}
\label{sec:ana_all}
\subsection{Phase averaged analysis}
\label{sec:isgri}

While data from individual observation blocks have previously been analyzed in detail
\citep{kreykenbohm08a, schanne07a}, studying all available data allowed us to
carry out a deep statistical analysis of the flaring behavior of
Vela~X-1. Our analysis
was performed with special regard to the giant flares with peak luminosities of more than six times the average luminosity \citep{staubert04a, kreykenbohm08a}, i.e., more than 300\,cps in ISGRI ($\sim$1.3\,Crab) between 20--60\,keV. It was not clear if these giant flares can be explained by
the same physical process as the other, common, flares in \mbox{Vela~X-1} or if they are caused by
a different mechanism. 

For this work, only ISGRI data between 20--60\,keV were analyzed, because in this energy range the source is bright and  the flux is unaffected by photoabsorption, because typical values of the equivalent hydrogen column $N_\text{H}$ outside the eclipse are below $3\times 10^{23}$\,cm$^{-2}$ \citep{kreykenbohm99a}.
Our analysis extends the energy range of prior observations of the stellar wind, which used high-resolution spectra in soft X-rays to investigate the
features of the wind \citep[and references therein]{watanabe06a}.

We extracted lightcurves from ISGRI 
using \texttt{ii\_light}, a tool distributed as part of the Offline Scientific Analysis (OSA) 7.0 package. The time resolution was chosen
to be 283.5\,sec to average each data-point over one pulse period to
eliminate these fluctuations. The pulse period is changing in a random-walk behavior on all times scales \citep{deeter89a}, but no long-term trend is evident, so that the chosen time resolution is sufficient for the purpose of this investigation.
\begin{figure}
   \centering
    \includegraphics[width=0.95\columnwidth]{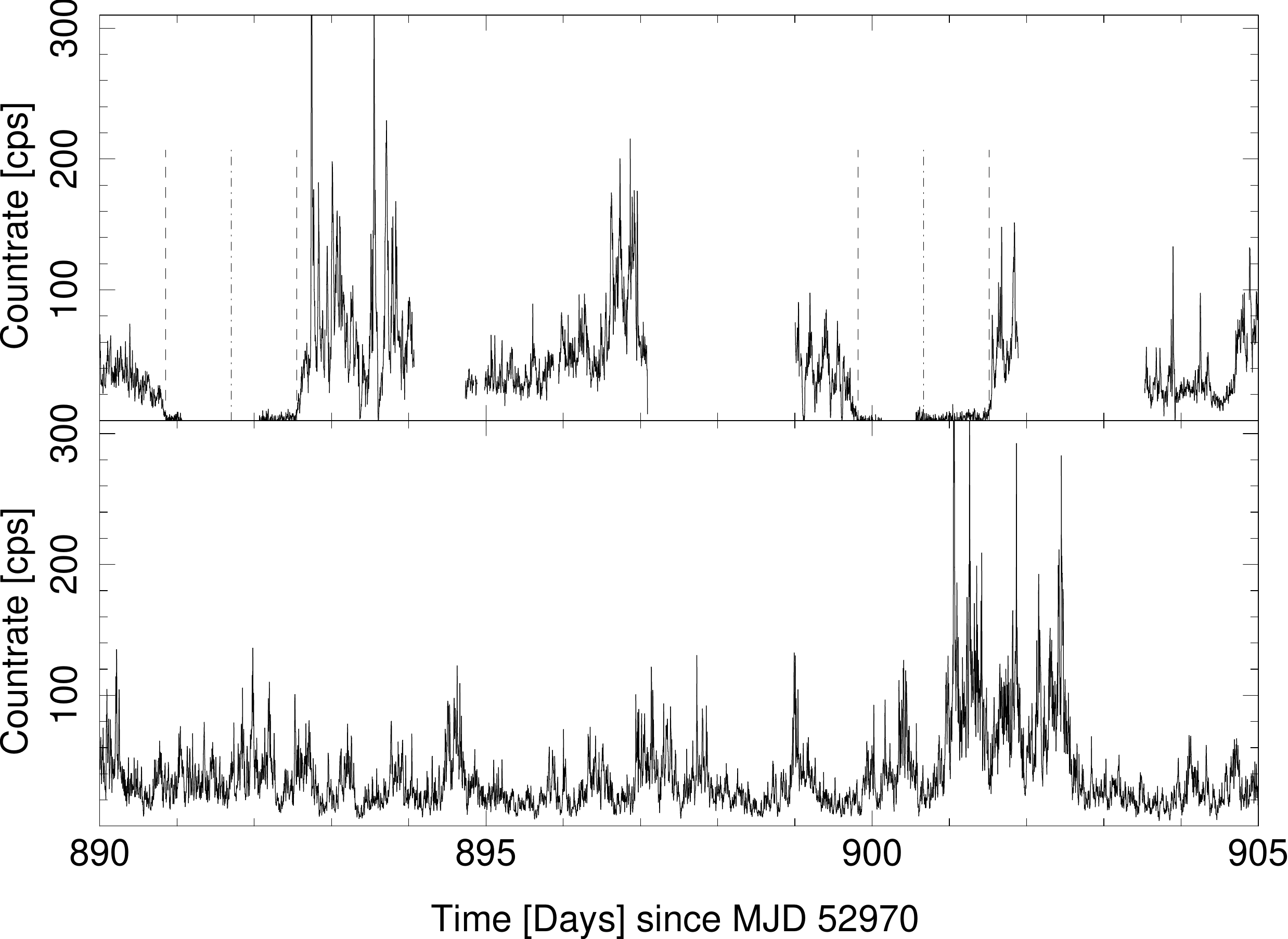}
   \caption[Lightcurve]{\textit{Top:} Lightcurve in the 20--60\,keV band. The vertical dashed lines show the start and the end
of the eclipse according to the ephemeris of \citet{kreykenbohm08a}, and
the dash-dotted line shows the respective center of eclipse.
\textit{Bottom:} Simulated lightcurve with the same
statistical parameters and temporal resolution as the observed lightcurve above, but without eclipses. See text for details. }
   \label{fig:lc_20-60_4xx.pdf}
 \end{figure}

As an example for the analyzed data the upper panel of
Fig.~\ref{fig:lc_20-60_4xx.pdf} shows the lightcurve of the 20--60\,keV band for
data
from revolutions 433--440 (Block 3).

\begin{figure}
   \centering
    \includegraphics[width=0.95\columnwidth]{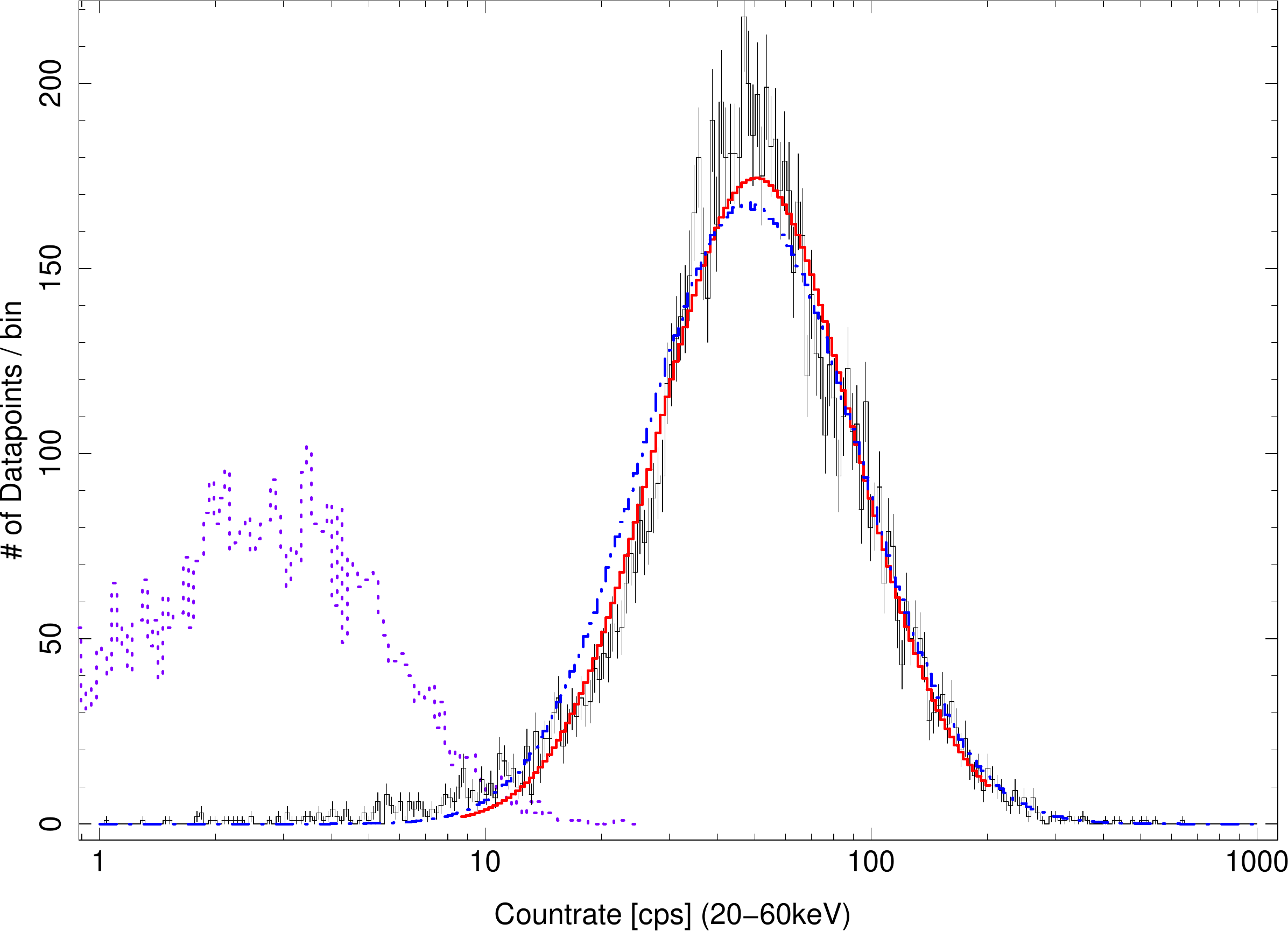}
   \caption[Histogram of LC in 20--60\,keV band]{Histogram of the lightcurve of
the 20--60\,keV band, binned to 256 bins. The solid curve shows the best-fit
single Gaussian, the dash-dotted one the histogram of the simulated lightcurve as
described in the text. The dotted histogram to the left shows the noise level of the background.}
   \label{fig:lcratehistall.ps}
 \end{figure}

The lightcurve data were filtered according to orbital phase $\phi_\text{orb}$, leaving only data of phases between 0.19 $\le \phi_\text{orb} \le$ 0.81, for which the system is out of eclipse. These data were then binned into 256 count rate bins. The bins are spaced logarithmically between 1\,cps and 1000\,cps.
 This binning leads to a histogram of the orbital phase averaged brightness distribution of the source, shown in black in
Fig.~\ref{fig:lcratehistall.ps}. The
distribution closely resembles a normal distribution in log-space, i.e., a log-normal distribution \citep{fuerst08b}.
In order to quantify the shape of the distribution we fitted a
Gaussian function in log-space, i.e., a log-normal distribution in countrate space. Following the approach by \citet{uttley05a} the uncertainties of the values $N_i$ of each histogram data bin were estimated to be $\sqrt{N_i}$, i.e., assuming Poisson statistics. We show that this assumption is justified in Sect.~\ref{sec:simu}.
The best-fit function, with a $\chi^2$-value of $190$ for 81 degrees of freedom, is shown as solid line in Fig.
\ref{fig:lcratehistall.ps}, where in the fit only bins with more than 20 measurements were taken into account \citep[see][]{gehrels86a}.
A Gaussian function in log-space is best characterized by its mean and standard deviation. These values represent the median of the log-normal function, which will be denoted as $\left<x\right>$, and its multiplicative standard deviation, $\tilde \sigma$, i.e., 68.3\% of all data points fall in the interval $[\left< x\right> / \tilde \sigma, \left< x\right> \cdot \tilde \sigma]$ \citep{ahrens54a}, resulting in asymmetric error bars according to the skewness of the distribution. Note that  consequently $\tilde \sigma$ is unitless.
With these definitions, the best-fit function has a median value of $\left< x\right>_\text{fit} =
50.3$\,cps and a multiplicative standard deviation of $\tilde{\sigma}_\text{fit} =
1.8$. 
Compared with that the measured count rates have a
median flux of $\left< x\right>_\text{meas} = 48.2$\,cps and a multiplicative standard deviation of $\tilde \sigma_\text{meas}=2.0 $, similar to the fit. 

The largest cause of source-independent variation that affects the data are
background fluctuations. To obtain  an estimate for the background noise level, we extracted a
lightcurve in the 20--60\,keV energy band from a region in the sky without any known X-ray
source ($\alpha = 9$h$\,16$m$\,37$s, $\delta = -44\degr 07\arcmin 11\arcsec$). The
brightness distribution of this background is shown with a dotted line in Fig.~\ref{fig:lcratehistall.ps}. The median value of the background distribution is
$\left<x\right>_\text{bkg}\approx$1.8\,cps, which is below 99.98\% of the source
data. Note that the brightest data points of the
background reach values of up to 11\,cps, which is in the lower left flank of the source
distribution. They do not, however, change the overall shape and parameters of
the distribution, so that a significant background influence can be ruled out.

\subsection{Orbital phase resolved analysis} 
\label{sec:phasres}

\begin{figure}
   \centering
   \includegraphics[width=0.95\columnwidth, viewport=0 0 449 445, clip]{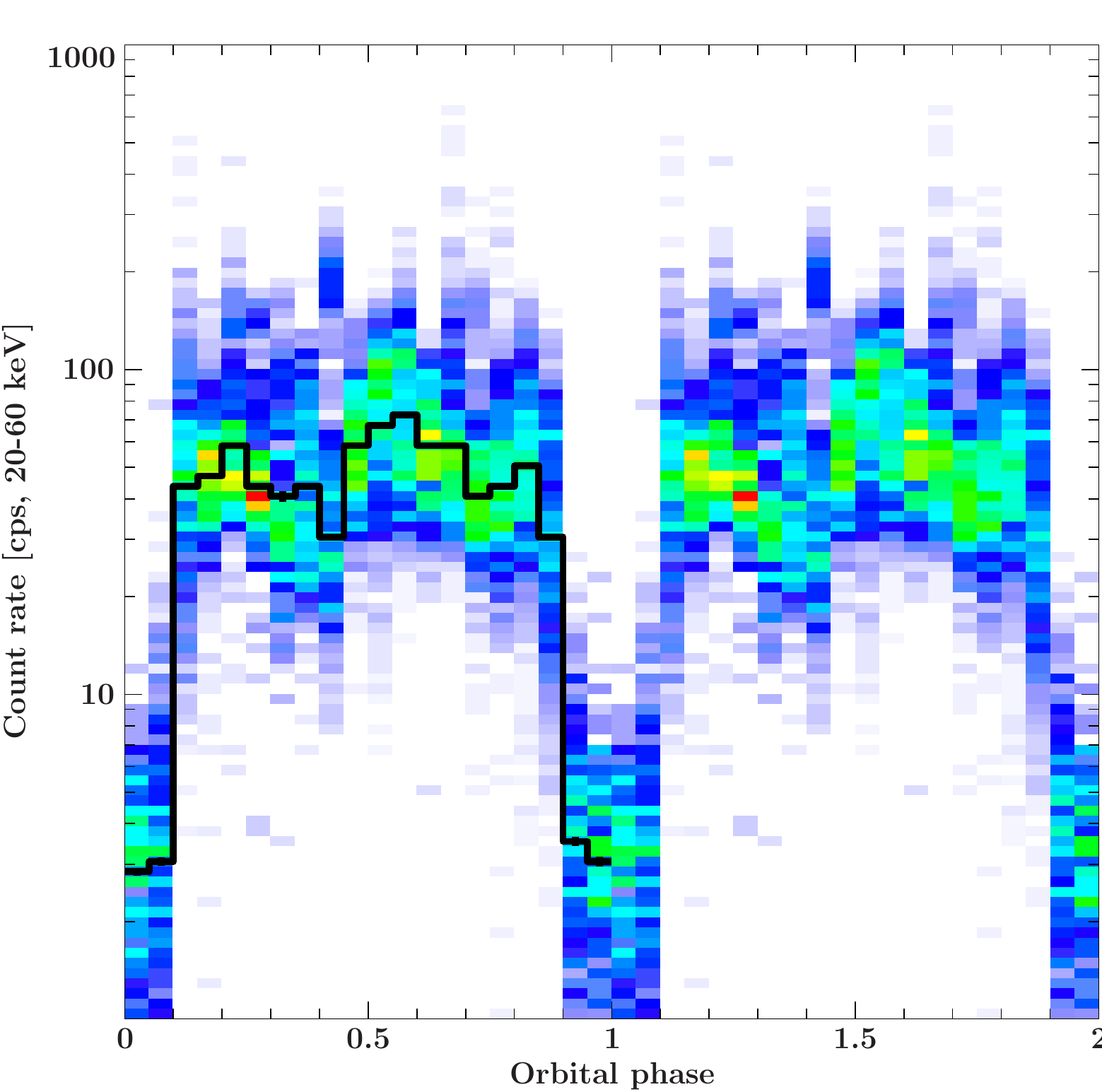}
   \caption{Landscape plot of orbital phase resolved histograms of the ISGRI 20--60\,keV data. Color-coded is the probability for a datapoint to fall into the respective histogram bin. The black line shows the median count rate in each phase bin. Uncertainties of that value are also plotted, but they are too small to be clearly visible in this plot. Note that the same histograms are shown twice for clarity. }
   \label{fig:phasres_histo.eps}
 \end{figure}

We have seen in the previous section that deviations from a log-normal distribution are visible. These deviations could be induced by averaging over all orbital phases, where possibly not all phases show the same statistical brightness variations. Systematic influences could be due to the complex and turbulent accretion geometry, consisting of a bow shock, an accretion wake, a possible photoionization wake or even a tidal stream \citep[see, e.g.,][]{blondin90a, blondin91a, mauche08a}. An accretion wake in Vela X-1 was already postulated by \citet{eadie75a}, showing up as absorption dips in the lightcurve of the Ariel V Sky Survey Experiment. 

To analyze the orbital influences, we extracted orbital phase resolved histograms. The data set comprises $\sim$8 orbits overall, which allows us to extract histograms with $\Delta \phi_\text{orbit} = 0.05$. We reduced the number of logarithmically spaced bins to 96 to collect more than 20 values in most bins to satisfy the requirements for $\chi^2$ statistics. Figure~\ref{fig:phasres_histo.eps} shows color-coded the probability of finding a datapoint in a countrate bin for a given histogram, equivalent to the height of the histogram in Fig.~\ref{fig:lcratehistall.ps}. Additionally the average countrate in every bin is shown, together with its uncertainty. As the average countrate can be calculated from a large number of data points in the lightcurve, these uncertainties are rather small. Significant variations from phase bin to phase bin can be seen in the figure. Especially notable is a dip in the average countrate around phase $\phi\approx0.3$--$0.4$ and a sudden brightening to twice the value afterwards. \citet{goldstein04a} analyzed \chandra data taken at $\phi_\text{orb}=0.25$ and $\phi_\text{orb}=0.5$ of one orbit and measured a stronger absorbed spectrum in the $\phi_\text{orb}=0.5$ data. These authors explained the increased absorption by the accretion wake being in the light of sight during that episode. The dip in Fig.~\ref{fig:phasres_histo.eps} occurs 
slightly before $\phi=0.5$. If an accretion wake is responsible for this feature, it seems to have shifted to earlier orbital phases in our data. 

When analyzing the three data blocks (Table~\ref{tab:blocks}) separately an orbital modulation is also visible, although with different strength in the different blocks. This indicates a strongly variable feature rather than an orbitally stable one and suggests that the variation seen in Fig. \ref{fig:phasres_histo.eps} is biased by single orbits probably deviating strongly from the average orbital histogram, maybe due to a particularly active or quiet state of the system.

The influence of different $N_\text{H}$ values during the orbit can be investigated by analyzing data in softer X-rays. Softer energy bands are far more affected by photoabsorption than the ISGRI band, because the cross-section of the photoabsorption is declining quickly with increasing energy as $\sigma \propto
E^{-3}$.  A very good data set in the soft X-rays is provided  by the All-Sky Monitor \citep[ASM,][]{levine96a} aboard the ``Rossi X-Ray Timing
Explorer'' (\xte). The ASM
performs 90\,sec dwells on an irregular basis onto different sources, including Vela~X-1, and thus provides good
statistics in the three energy bands between 1.5--10\,keV called A (1.5--3\,keV), B (3--5\,keV), and C (5--10\,keV) band. 

We performed a similar analysis of the
ASM data as of the ISGRI data, but binning the countrate directly to
256 bins instead of its logarithm. The linear binning was necessary because especially in the eclipse many data points of the lightcurve are negative and cannot be presented on a logarithmic scale.
It has previously been shown that due to the complex accretion geometry of the system, the absorption coefficient frequently changes dramatically over the orbit \citep{haberl90a, kretschmar08a}. Especially in the late orbital phases after $\phi_\text{orb} = 0.5$, the accretion and photoionization wake, consisting of denser material, are expected to be in the line of sight \citep{kaper94a}. 
To analyze the influence of the absorption on the brightness distribution, we divided the data into the same 20 orbital phases as the ISGRI data. 

\begin{figure}
   \centering
    \includegraphics[width=0.95\columnwidth, viewport=0 0 443 678, clip]{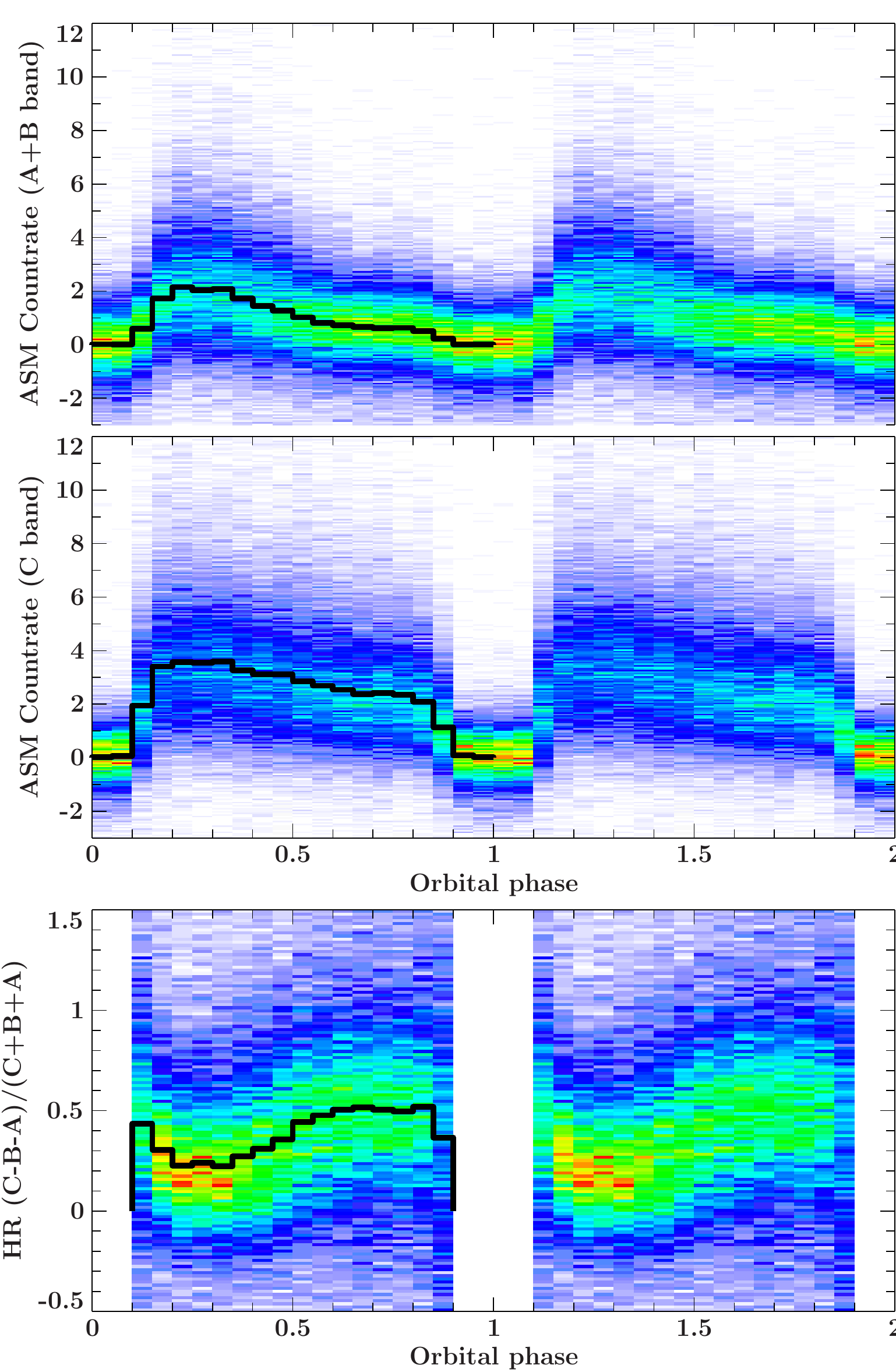}
   \caption[Histogram of LC of the ASM in the 2--10\,keV band]{Same as Fig. \ref{fig:phasres_histo.eps}, but for the ASM data in the 1.5--5\,keV (top) and 5--10\,keV band (middle). The bottom panel shows the hardness ratio, overplotted is the median of the distribution. Data in the eclipse were ignored. A clear trend to a harder spectrum at late orbital phases is visible.}
   \label{fig:ctratehisto_asm.eps}
 \end{figure}

Figure~\ref{fig:ctratehisto_asm.eps} shows the histograms for the A+B band and for the C band. We added the narrow A and B band as both showed very similar behavior, but only low significance. By adding A and B the countrates are comparable to the C band. It is obvious that in both bands the mean luminosity is monotonically decreasing over the orbit \citep[see also][]{wen06a}. At the same time, the distribution narrows, meaning that strong deviations from the mean are more common in the early orbital phases than in the later ones. It is remarkable that no sudden dip structures like in the ISGRI data in Fig. \ref{fig:phasres_histo.eps} are visible. This could be because the ASM data stretch over more than 13\,years, averaging out any individual orbit variations. As the ISGRI data comprise only eight orbits, they are strongly influenced by single outliers and do not provide a sufficient time basis for determining an average orbital profile. Nonetheless, no declining trend in the average brightness is evident in ISGRI. 

The missing trend strongly points to the explanation that the trend in the ASM is due to increasing absorption by a cold absorber. This interpretation is supported by the analysis of the hardness ratio, which is increasing with orbital phase (Fig. \ref{fig:ctratehisto_asm.eps}, bottom panel). The first phase bin after eclipse is also significantly harder on average, because the X-ray source is still behind the extended stellar atmosphere of the optical companion \citep{kretschmar08a}. In contrast to the countrate histograms, the hardness ratio shows a larger variation at later orbital phases. This broadening again indicates that no persistent features are present in the wind, but that any absorbing material is only temporary and that the column density is varying from orbit to orbit.

Looking at the distribution of the ASM as data, we find that they follow a log-normal distribution only coarsely.
We ascribe this to the strong influence of photoabsorption on the ASM lightcurve, reducing the apparent brightness. This influence means that the observed distribution is not only caused by flaring, i.e., changes in the mass accretion rate, but also by varying \nh. This entanglement makes it impossible to infer the distribution of the accreted masses from the measured distribution.

\section{On generating log-normal distributions}
\label{sec:simu}
In the previous sections we have seen that the brightness distribution above 20\,keV is nearly log-normal. In the peak of the distribution relatively large deviations are, however, visible and the overall reduced $\chi^2$-value of the fit is still high. All attempts at fitting a log-normal distribution to the rising flank up to the peak resulted in unacceptable
$\chi^2$-values only. These deviations from the log-normal distribution are a hint that the underlying processes are not purely multiplicative \citep{uttley05a}. 
As the log-normal distribution is very well understood we demonstrate in this section a method to simulate a lightcurve showing a log-normal brightness distribution. These simulations illustrate the physical processes that are at work in a system like Vela~X-1. Furthermore they help us to estimate the uncertainties in the distributions and to separate the influence of the sampling from intrinsic source behavior.
To simulate the lightcurves we implemented the method proposed by \citet{uttley05a}, introducing slight modifications to allow other mean values than 0.

The method used is based on the fact that any linear
 lightcurve $L(t)$ can be written according to the Fourier theorem as a linear superposition of individual sinusoids with fixed frequencies $\nu_i$ and normalizations $A_i$ and phases $\phi_i$, i.e.
\begin{equation}
 L(t) =  \sum_{i=1}^\infty A_i \sin(2\pi \nu_i t + \phi).
\end{equation}
On the other hand the Fourier transform of a lightcurve can also be represented by a superposition of individual sinusoids. By drawing the amplitudes and phases of these sinusoids from a normal distribution and calculating the inverse Fourier transformation of these values, a linear, stochastic lightcurve can easily be simulated \citep{timmer95a}. To obtain a simulated lightcurve resembling observational data, randomly drawn amplitudes and phases of each frequency are weighted by a filter function that describes the frequency spectrum of the observed data. This filter function is given by the shape of the periodogram of the data, as only real values are important for the lightcurve \citep{uttley05a}.
 According to the central limit theorem, the simulated lightcurve will have a normal brightness distribution, because it can be written as the sum of random numbers drawn from the same probability distribution. 
\citet{uttley05a} pointed out that in much the same way, a lightcurve with a log-normal brightness distribution can be simulated by using the product instead of the sum of randomized sinusoids. This is most easily achieved by taking the exponential function of a linear lightcurve. \citet{uttley05a} used the method by \citet{timmer95a} to obtain the linear lightcurve, an approach we also adopted for our analysis.

\begin{figure}
   \centering
    \includegraphics[width=0.95\columnwidth]{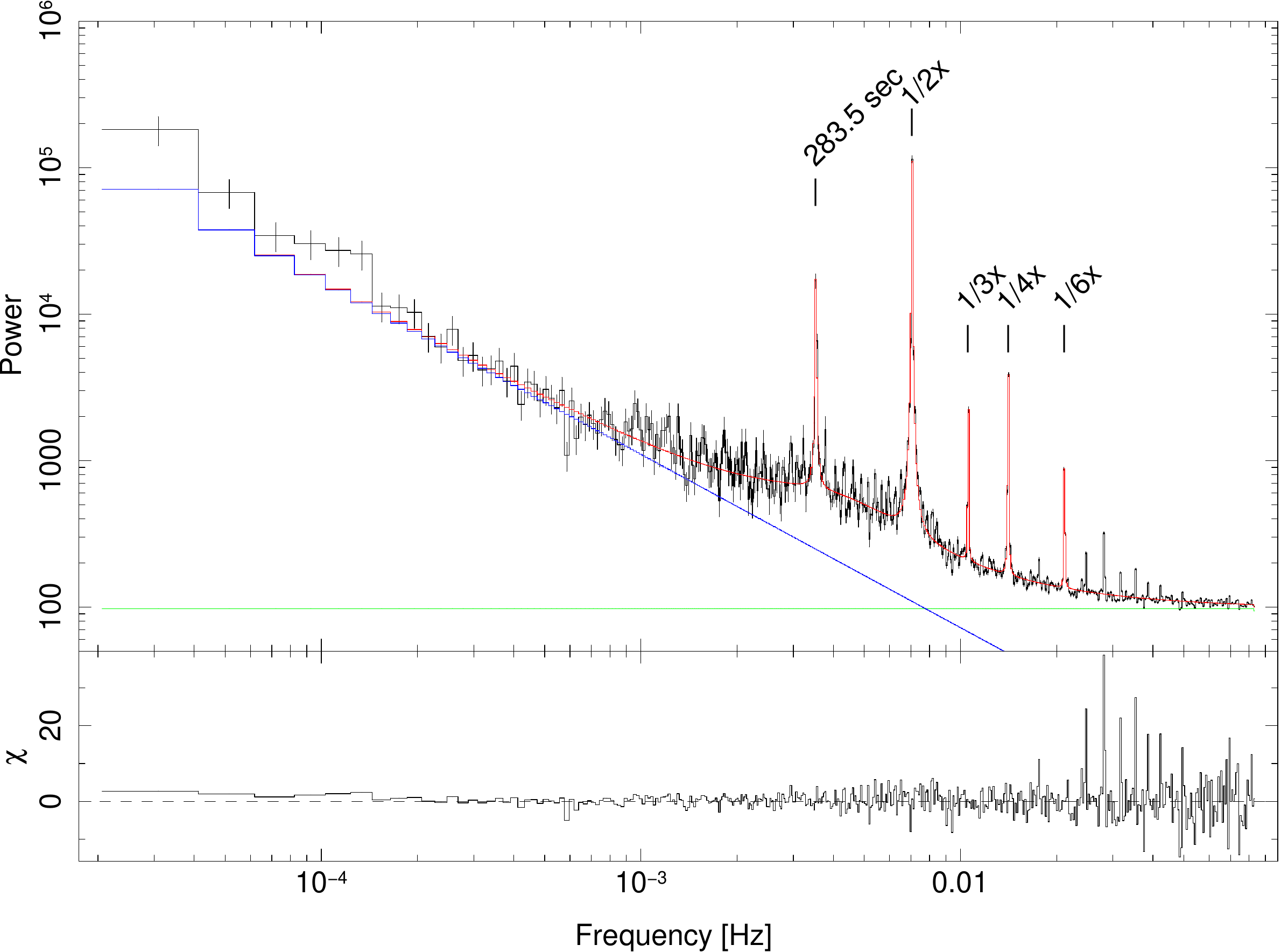}
  \caption[PSD]{Showcase PSD of Vela~X-1 from Revs 373--383 in the 20--100\,keV ISGRI band. The red curve shows the best-fit model, as described in the text. The blue power law continuum has a $\gamma = 1.19$. The peaks of multiples of the pulse frequency were modeled using Lorentzians, with the center of the Lorentzian fixed to the respective multiple of the pulse frequency (see Table \ref{tab:psdmodel}).}
   \label{fig:psd_fit.eps}
 \end{figure}

In order to evaluate the shape of the power spectral density (PSD) of Vela~X-1 we extracted a lightcurve with 6\,sec resolution in the 20--100\,keV energy range and divided it into segments of evenly spaced data of 49152\,sec ($2^{13}\times6$\,sec). In order to obtain evenly spaced lightcurves, we interpolated over small gaps around the slew time scale of $\sim$150\,sec. This interpolation reduces spurious power on the ScW timescale. The average PSD for all segments drawn from data of Rev. 373--383 (Block~2) is shown in Fig. \ref{fig:psd_fit.eps}, together with the best-fit model consisting of a constant plus a power law and 6 Lorentzian lines. The Lorentzians are characterized by their peak frequency, their width and their normalization, describing the overall contribution to the PSD. They are necessary to model the contribution of the strong pulse with a pulse period of 283.5\,sec and its harmonics, as marked in Fig. \ref{fig:psd_fit.eps}. The frequencies of the Lorentzians were fixed to multiples of the pulse frequency. The first two Lorentzians are both located at the pulse frequency, with Lorentz \# 1 modeling a broad base, while Lorentz \# 2 models the narrow spike. The inclusion of the broad component drastically improves our fit. A similar superposition of two Lorentzians at the pulse frequency was also seen in V~0332$+$53 by \citet{mowlavi06a}, who  interpreted it as a quasi periodic oscillation (QPO). A detailed analysis of this feature in Vela X-1 is, however, beyond the scope of this paper.  

It is interesting to note that the power at half the pulse period ($7.05\times10^{-3}$\,Hz) is more than 4 times higher than at the actual pulse period. This effect is caused by the pulse shape in the regarded spectral energy range consisting of two very similar pulses, which could easily be misunderstood as pulsations with half the actual pulse period \citep{kreykenbohm99a}. 
The underlying continuum of the PSD can be modeled accurately with the constant and the power law component of the model, with the power law having the shape $\nu^{-\gamma}$, where $\gamma = 1.19$. All fit parameters are listed in Table \ref{tab:psdmodel}. 

The noise level of \inte ISGRI PSDs is poorly understood, but a comparison with \xte PCA PSDs shows that for instrumental reasons it is not purely Poissonian. We are currently in the process of analyzing other \inte data of other HMXB to investigate ISGRI's broad band timing capabilities. For the present analysis, however, only the slope of the power law of the PSD is important, which is independent of the overall noise level. We compared the ISGRI result to PCA PSDs of Vela~X-1 and found a value of $\gamma \sim 1.4$ for the latter. The difference does not change the outcome of the simulation.
Contributions from the pulse period to the PSD can 
also be neglected, as the sampling rate of the lightcurve is chosen to eliminate them.

\begin{table}
\centering
\caption{Fit parameters of the PSD (Fig. \ref{fig:psd_fit.eps})}
\begin{tabular}{ll}
\hline\hline
  Component & Value \\\hline
 Constant  &  $94.67^{+0.51}_{-0.54} $ \\
Powerlaw (Norm)  & $0.296^{+0.033}_{-0.032}$ \\
Powerlaw (Gamma) &  $1.191 \pm {0.019} $  \\
Lorentz 1 (Norm) & $2.70^{+0.16}_{-0.17}$ \\
Lorentz 1 (Freq) (Hz) & $1/283.5$  \\
Lorentz 1 (Width) (Hz) & $(0.57\pm0.03)\times10^{-2}$   \\
Lorentz 2 (Norm) & $0.90 \pm 0.13 $  \\
Lorentz 2 (Freq) (Hz) & $1/283.5$  \\
Lorentz 2 (Width) (Hz) &  $(2.09^{+0.49}_{-0.38})\times10^{-5}$	  \\
Lorentz 3 (Norm) &  $10.76^{+0.84}_{-1.44}$ \\
Lorentz 3 (Freq) (Hz) & $2/283.5$ \\
Lorentz 3 (Width) (Hz) &  $(1.91^{0.33}_{-0.15})\times10^{-5}$  \\
Lorentz 4 (Norm) &   $0.254 \pm 0.020$ \\
Lorentz 4 (Freq) (Hz) & 3/283.5  \\
Lorentz 4 (Width) (Hz) &   $(1.51^{+0.32}_{-0.28})\times10^{-5}$  \\
Lorentz 5 (Norm) & $0.642 \pm 0.033$  \\
Lorentz 5 (Freq) (Hz) & 4/283.5  \\
Lorentz 5 (Width) (Hz) &   $(2.47^{+0.22}_{-0.20})\times10^{-5}$ \\
Lorentz 6 (Norm) &  $(20.09 \pm 0.82)\times10^{-2} $ \\
Lorentz 6 (Freq) (Hz) & 6/283.5  \\
Lorentz 6 (Width) (Hz) &   $(2.00_{-0.20}^{+0.21})\times10^{-5}$  \\\hline
\end{tabular} 
\label{tab:psdmodel}
\end{table} 

The
simulation was then carried out using a PSD  of the shape $\nu^{-1.19}$ and taking care that
the emerging statistical parameters $\left<x\right>_\text{sim}$ and $\tilde \sigma_\text{sim}$  are
the same as for the input data. A part of the simulated lightcurve is shown in
the lower panel of Fig. \ref{fig:lc_20-60_4xx.pdf}. The overall behavior of the data, showing short, strong flares can
be very well described with the model.
The PSD has a relatively large power at low frequencies,
showing up in notable variations in the statistical parameters of the
emerging lightcurve between individual runs of the simulation. To determine
the size of these variations, we simulated 250 lightcurves with the same
parameters $\left<x\right>$ and $\tilde\sigma$. As the simulated lightcurve incorporates a frequency spectrum of similar shape as the data, the uncertainty levels of the simulation can be used to estimate the uncertainty of each histogram bin of the data. 
The standard deviation in the simulation in each bin was on the order of the Poisson statistic, justifying the Poisson approximation used in Sect. \ref{sec:ana_all}. The distribution of
the simulated lightcurve describes the measured histogram almost as  well as the
best-fit Gaussian ($\chi^2 = 216$ in 84 bins, see Fig. \ref{fig:lcratehistall.ps}). 

\section{Discussion and conclusion}
\label{sec:disc}
\subsection{Summary}
We presented the first analysis of the statistical distribution of the
brightness of Vela~X-1, with special regard to the flaring behavior. Our main results are:
\begin{itemize}
 \item The orbital phase averaged brightness distribution above 20\,keV can be characterized by a log-normal distribution.
 \item The orbital phase resolved histograms indicate systematic variations besides the eclipse.
 \item The brightness distribution in the soft X-rays is strongly affected by absorption.
 \item Analysis of the soft X-rays confirm a decline in the average countrate with orbital phase.
 \item The brightness distribution of a simulated lightcurve consisting of a multiplication of individual random processes  can be used to describe the data.
\end{itemize}

\subsection{Giant flares}
From our results obtained by analyzing 3.6\,Msec of \inte ISGRI data we can calculate the probability to measure extremely bright flares in that time range. Extremely bright flares are defined by their countrate being larger than
300\,counts\,sec$^{-1}$ in the 20--60\,keV band. The calculated probability is $\sim$0.011\% for the best-fit Gaussian, while it is $\sim$0.023\% for the simulated lightcurve. These numbers agree well with $\sim$0.02\%  measured directly from our dataset.
We conclude that bright flares are rare, but not singular events. No
separate process is required for their explanation. 

\subsection{Wind structure and accretion geometry}
The origin of the constant flaring behavior lies in the accretion flow onto the neutron star, which is obviously not smooth, but highly structured. 
Neglecting any absorption and scattering effects, the luminosity distribution gives us the opportunity of calculating the mass distribution in this flow in units of mass per time. To calculate the mass accretion rate it is necessary to know the absolute luminosity of the X-ray source in the given energy band. As the luminosity is strongly energy-dependent, we modeled the energy spectrum with a power law with Fermi-Dirac-cutoff, using the parameters of
\citet{kreykenbohm08a}, who found these parameters for data from revolutions 137--141, 
which were part of our analysis. The overall spectrum of our data could be described with these values, too.
With those parameters and a distance of 2\,kpc \citep{nagase86a} the median absolute luminosity $\left<L_\text{x}\right>$ is 
$ 5.1\times10^{36}$\,erg\,sec$^{-1} $.

\begin{figure}
   \centering
    \includegraphics[width=0.95\columnwidth]{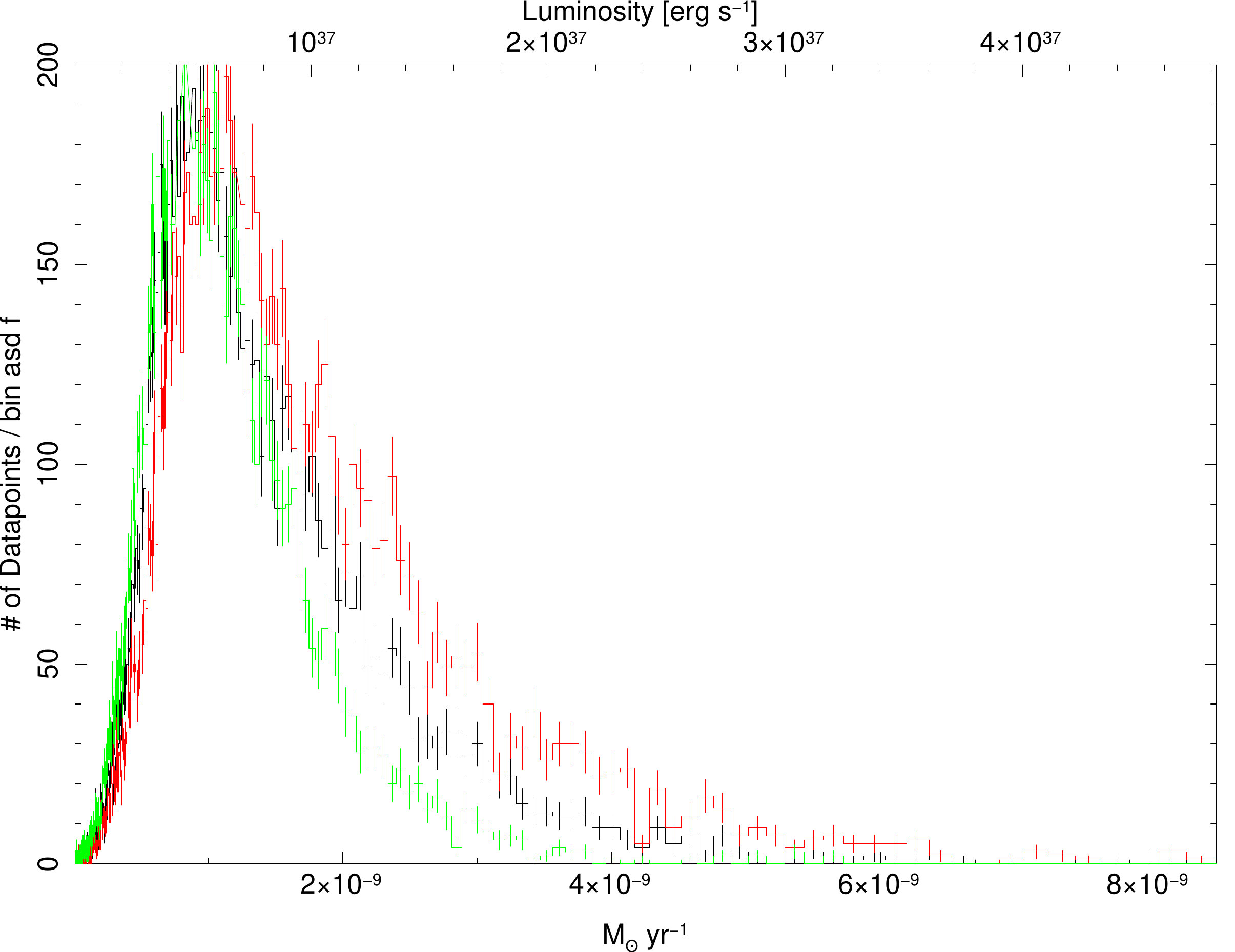}
   \caption[Histogram of mass accretion rate]{Histogram of the calculated
accretion rate $\dot{M}$ of the neutron star. The rate was calculated using a
standard spectrum and the countrate in the 20--60\,keV band (black), 20--30\,keV
band (red), and 40--60\,keV band (green). For detailed description see text. 
}
   \label{fig:histo_mdost.eps}
 \end{figure}

During the accretion process only part of the potential energy is converted to X-rays. We assume an accretion efficiency of $\eta = 0.3$ and calculate the median accretion rate 
\begin{equation}\left<\dot M\right> =  \frac  {\left<L_\text{x}\right>}  {c^2 \cdot \eta} = 3.0\times10^{-10}\,\text{M}_\odot\,\text{yr}^{-1} 
\label{eq:mdot}
\end{equation}
The multiplicative standard deviation is $\tilde \sigma_{\dot M}  = 1.9$ in the 20--60\,keV energy range. Figure \ref{fig:histo_mdost.eps} shows the distribution of inferred $\dot M$ values and the corresponding luminosity. To show the log-normal behavior of the distribution more clearly, we used a linear scaled $x$-axis. We included data not only from the 20--60\,keV energy band, but also from the softer 20--30\,keV band and the harder 40--60\,keV band in this figure. All three bands show the same behavior. Slight deviations between the different energy bands are due to the rough spectral fit, because only the limited energy range of ISGRI was available for fitting. Nonetheless it is evident that all three
curves follow a log-normal distribution in a range of $\dot{M}$ very well
expected for this kind of object regarding the mass loss rate of the optical
companion and the assumed accretion efficiency. 

\subsubsection{A clumpy wind}
If we assume that the neutron stars accretes directly from the wind, we can infer the mass distribution in the stellar wind from the accretion rate. Strong density variations in the wind comparable to the variations in the accretion rate can be described by a model of a clumped stellar wind, a model based on instabilities in the the line-driven acceleration mechanism 
\citep[see, e.g.,][]{feldmeier03a, dessart05a, oskinova07a}. A clumpy wind is also supported by observations. Based on \textsl{ASCA} data, \citet{sako99a} show that the
stellar wind of HD\,77581 is most likely strongly clumped by measuring line strengths in the spectra and showing that these lines are only in accordance with a mass loss rate of $\sim$$10^{-6}\,M_\odot\,$yr$^{-1}$ when assuming a
clumped wind, consisting of dense, cold clouds and a hot, ionized surrounding
medium. The clouds make up for about 90\% of the wind mass, while the hot
ionized medium makes up for about 95\% of the volume, i.e., the volume filling factor of the clumps is $f_v=0.05$. \citep{sako03a}.  
Further evidence for clumping in the stellar wind was brought forth by different authors using high-resolution \chandra spectra \citep{schulz02a, goldstein04a, watanabe06a}. In these spectra emission lines from H- and He-like ions indicate a hot thin gas. The spectra, however, showed additionally strong fluorescent lines from lower ionized atoms, e.g, of Fe, Mg, and Ne. These originate in colder, optically thick plasma, which can be explained as clumps in the hot surrounding gas. 
Additionally, if a dense clump is passing through the line of
sight between the observer and the X-ray emitting region, the absorption
coefficient will rise significantly.  Clumps measured
through absorption dips \citep{nagase86a} may very well be the same clumps that are accreted and
are thus directly responsible for luminosity fluctuations measurable up to at least
100\,keV \citep{sako99a}.

The accretion of a clump or a similar dense feature in the wind will lead to an increased X-ray luminosity of the source, i.e., a flare. As the lightcurve of Vela~X-1 is dominated by those flares, which are strongly overlapping, it is hard to estimate the length of any individual flare. Visual inspection of the lightcurve indicates, however, that the shortest flares last around $t_\mathrm{fl} \approx 2.5$\,ksec. During that time $\langle M \rangle_\text{acc} = \langle\dot{M}\rangle t_\mathrm{fl} = 4.7\times10^{19}$\,g of material will be accreted onto the neutron star, assuming the average accretion ratio from Eq.~\ref{eq:mdot}. 
Using the $\beta$-law \citep{castor75a}, the average density at the neutron star orbit gives values of $\rho_\mathrm{wind} \approx 8\times10^{-18}\,\mathrm{g}\,\mathrm{cm}^{-3}$. In their simulations \citet{blondin90a,blondin91a} showed that regions with an overdensity of $\sim$100 are possible in the filaments of the accretion wake. Assuming a spherical clump with such an increased density, i.e., a volume filling factor $f_v=0.01$ \citep{fullerton06a}, a radius of $r_\mathrm{cl} = 2.37\times10^{10}$\,cm is obtained. These clumps would show up in absorption with a maximal column density of $N_\mathrm{H}^\mathrm{cl} \approx 9\times 10^{21}\,\mathrm{cm}^{-2}$, i.e., close to the observed absorption column \citep{kreykenbohm08a}, especially considering that more than one clump could be in the line of sight at any given moment and that we assumed short flares, and therefore small clumps.

Besides the  normal flares, giant flares are also visible in the lightcurve and are more easily constrained in time. The largest flare in the
investigated lightcurve is at MJD 52971.2, lasting for 38\,ksec with an average count rate of $\sim$140\,cps. During that time  $M_\text{gf} \approx 2\times 10^{21}$\,g were accreted, using the same conversion factor and efficiency as in Eq.~\ref{eq:mdot}. 
This mass is in the regime of
the values given in numerous other works, see for example \citet{negueruela08a} or
\citet{walter07a}. Note, however, that these authors have
investigated supergiant fast X-ray transients (SFXT), a rather new class of
X-ray sources. In a model proposed by \citet{negueruela06a}, these SFXT
are regarded as HMXB, similar to systems like Vela~X-1, with the orbit
parameters being the major difference. This similarity extends to similar companion stars, and thus similar wind properties.
The similar wind properties are also evident in the irregular flaring behavior of the SFXT IGR J08408$-$4503, which was explained by a clumpy wind \citep{leyder07a}. This explanation was also given for the flaring behavior of ``regular'' HMXB \citep[see, e.g.,][]{vandermeer05a} and is supported by our analysis. 
We conclude that the calculated clumpsizes and masses for SFXT can be very well compared with the values of large clumps of Vela~X-1. 

\subsubsection{Temporal smearing}
In the previous section we showed that the clump distribution is probably roughly log-normal. If true, this result contradicts to typical assumptions for mass distributions in the atmospheres of high-mass stars, where power-law distributions are generally assumed \citep[e.g.,][]{kostenko99a}. 
But it is unlikely that the
distribution of clumps in the stellar atmosphere translates directly to a
distribution of accretion rate from the wind, because temporal smearing is taking place during accretion. 
If the clump mass distribution in the wind would be indeed log-normal, smearing could lead to small deviations from this log-normal distribution in the measured brightness distribution.
We tried to analyze the effect of smearing by folding the simulated lightcurve with an exponential function
with a characteristic timescale. Due to the log-normal character of the
lightcurve, strong smearing leads to a higher average countrate, which
disagrees with the measured lightcurve distribution. We therefore included
another fit parameter, allowing the renormalization of the simulated lightcurve. With this
method we could find no reliable timescale for the smearing, because our
time resolution was limited to 283.5\,sec and all emerging fit values were of
the order of or lower than that resolution. The analysis of a lightcurve with higher
time resolution of 20\,sec, but with a moving average of 283.5\,sec applied to
smooth out the intrinsic pulse variations came to the same result: the smearing
timescale is around the level of the time resolution and thus not significant.
From this simulation we can conclude that smearing cannot account for a measurable amount of deviation from the initial distribution of the clump sizes. From the observer's point of view we note that Vela~X-1 shows no evidence for an accretion disk, which would show up in the soft X-rays as a blackbody component. The spin period evolution also shows no evidence for an accretion disk, as the pulse period is changing erratically rather than with a constant trend,  which would be expected from the continuous transform of angular momentum from an accretion disk \citep{ghosh79a, deeter89a}. This lack of an accretion disk supports the argument of a short smearing timescale, because the accretion timescale is also short.  

\subsubsection{Shock structures and grinding}
We did not find in the literature any discussion covering the distribution of clump masses in these systems. Assuming that the clumps are log-normal distributed is therefore at the moment an ad hoc assumption and not explained by the models. In this section we will describe a possible mechanism to generate log-normal distributions. To do this, we will first take a closer look at the different features in the accretion region.

 Simulations by \citet{blondin90a, blondin91a} and recently by \citet{mauche08a} have shown that a shock front forms in front of the neutron star, together with an accretion wake and a possible photoionization wake, as the neutron star ploughs with supersonic motion through the stellar wind of the companion star.
These features induce additional variability of the X-ray flux over the orbit, an effect we have seen in Sect. \ref{sec:phasres}. Even though photoabsorption is negligible at the energies under consideration, Thomson scattering is not. The Thomson cross section depends only weakly on the photon energy, and Thomson scattering can scatter a considerable amount of X-rays out of the line of sight. As shown by \citet{hanke08a}, deep scattering troughs can be measured in \inte data of Cygnus X-1, which are ascribed to a large, hot clump passing through the line of sight. 
From the variations of the median of the orbital-phase-resolved histograms 
 we can calculate the necessary electron column to induce these changes and, assuming ISM abundances, the corresponding \nh value.
 The values obtained were on the order of $\nh=\left(20-120\right)\times10^{22}$\,cm$^{-2}$, depending on the energy band used.  Compared to that, the average \nh expected from the stellar wind is $\nh^{\text{Wind}} \approx 2.8\times10^{22}$\,cm$^{-2}$, assuming a wind density profile following a $\beta$-law \citep{castor75a} with $\beta = 0.8$.  This result is confirmed by measurements of the X-ray spectrum, which also give values in the regime of 1--$30\times10^{22}$\,cm$^{-2}$ \citep[e.g.,][]{kreykenbohm99a}. Assuming that the variations in our data are solely caused by scattering, the column density must increase up to a factor $\sim$50 compared to the average wind.  In simulations by \citet{blondin90a, blondin91a} regions with up to 100 times the density of the ambient wind are seen, which would lead to clumps with radii $r\approx10^{12}$\,cm and large masses on the order of $10^{24}$\,g, assuming spherical clumps. The donor star itself has only a radius of $R_\mathrm{star} = 2.1\times10^{12}$\,cm, making those clumps unrealistically large. The observed variations are thus unlikely to be induced by scattering alone.

The powerful structures in the accretion region will not only change the column density, but will also influence the mass distribution itself. All mass has to pass at least one shock wave and will be subject to strong turbulence prior to accretion. \citet{blondin91a} have shown that Rayleigh-Taylor instabilities will rise inside the bow shock region, producing overdense eddies and vortices, e.g., structures similar to clumps. It is also very probable that the mass distribution will be changed during this shock transition, as clumps can not only be produced, but large clumps of the stellar wind will be broken up into smaller pieces. Breaking up large clumps is a multiplicative process, and thus forms a log-normal distribution, which is long known in geoscience for the distribution of rocks and sand grains \citep[and references therein]{smith64a}. \citet{kevlahan09a} have shown that shocks in the interstellar medium can lead to a log-normal density distribution as well. 

Shock fronts and turbulence breaking up clumps can transfer any given distribution into a log-normal like distribution. The emerging masses of the clumps are uncertain however and need more and detailed simulations. A clumpy wind, with huge clumps with masses of at least $10^{22}$\,g, which is subject to shocks and turbulence grinding down these huge clumps to masses of $\sim$$10^{20}$\,g, could therefore be a realistic picture of the accretion geometry.  The clump distribution in the wind, however, would be transformed to a log-normal distribution of the accreted mass by means of the accretion processes, which we can see in the X-ray luminosity.

\subsubsection{Outlook}
Our analysis allows us for the first time to investigate the
structure of the wind in an energy band above 20\,keV. This structure has not been
investigated so far in this energy band in systems like Vela~X-1, but shows similar results of a strongly
structured wind as other investigations \citep[e.g.,][]{sako99a}. It is remarkable that the brightness distribution of the black hole HMXB Cygnus~X-1 shows a similar behavior. Although these systems produce their radiation
through different processes, the emerging distribution is also log-normal, as
shown for short time-scales by \citet{uttley01a} and \citet{gleissner04a}.  \citet{poutanen08a} were able to extend this result to longer time-scales.
While our results improve our understanding of the physical processes behind the
variations of the luminosity and the structure of the accreted mass, they do not allow to infer where the structure originates from, i.e., from clumps in the stellar wind or from effects during the accretion process. Further investigations should include more detailed models of the brightness distribution
and extensive magneto-hydrodynamic simulations of the accretion geometry and the influence of the neutron star on the wind. 
We propose that a strongly structured wind, with clumps of the masses around $10^{20}$\,g is accreted onto the neutron star, and thereby subject to shocks and turbulence. These disturbances can lead to a log-normal distribution in accreted masses, which translates directly to the observed luminosity distribution.

In addition, a similar analysis should be performed for other systems, providing more information about that class of
objects and their possible connection to SFXT. For example, a preliminary look at
4U 1909+07 \citep[$\equiv$ X1908+07,][]{wen00a} has shown that the brightness
distribution of this system is following a roughly log-normal distribution as
well.

\acknowledgements This work was supported by the Bundesministerium f\"ur Wirtschaft und Technologie through DLR grant 50\,OR\,0808 and via a DAAD fellowship. This work has been partially funded by the European Commission under the 7th Framework Program under contract ITN\,215212. FF thanks the colleagues at UCSD and GSFC for their hospitality. We thank M. A. Nowak and A. Pollock for the useful discussions. For this work we used the ISIS software package provided by MIT. We especially like to thank J. C. Houck and J. E. Davis for their restless work to improve ISIS and S-Lang. This research has made use of NASA's Astrophysics Data System. This work is based on observations with \inte, an ESA project with instruments and science data centre funded by ESA member states (especially the PI countries: Denmark, France, Germany, Italy, Switzerland, Spain), Czech Republic and Poland, and with the participation of Russia and the USA. We used the archival data from the ASM Light Curve web page, developed by the ASM team at the Kavli Institute for Astrophysics and Space Research at the Massachusetts Institute of Technology.
We thank the anonymous referee for her/his useful comments.

\bibliographystyle{jwaabib}

\end{document}